\documentclass[cameraready]{Interspeech}

\title{A Sensitivity Analysis of Multi-Event Audio Grounding in Audio LLMs}

\author[affiliation={1}, orcid=0009-0009-3576-2770]{Taehan}{Lee}
\author[affiliation={1}, orcid=0009-0001-6348-5397]{Jaehan}{Jung}
\author[affiliation={1}, orcid=0000-0003-2981-0800, correspondingauthor]{Hyukjun}{Lee}

\address{
    $^1$ Sogang University, South Korea
}

\email{alpaca@sogang.ac.kr, jhjung22@sogang.ac.kr, hyukjunl@sogang.ac.kr}


\usepackage{comment}
\usepackage{xcolor}
\usepackage{graphicx}
\usepackage{subcaption}

\begin{document}

\maketitle

\begin{abstract}
Audio LLMs have shown a strong ability to understand audio samples, yet their reliability in complex acoustic scenes remains under-explored.
Unlike prior work limited to small scale or less controlled query construction, we present a large-scale evaluation of event grounding and false alarms as auditory scene complexity increases.
Using 71K AudioCapsV2 clips, we extract normalized (source, attribute) events and build two query types:
present-event queries for ground-truth detection and absent-event queries to probe hallucinations, using similarity-filtered negative sampling in an audio-aligned text embedding space.
We evaluate four SOTA Audio LLMs with 12 prompt variants over 500K yes/no queries per model.
Across models, increasing event count consistently lowers true-positive rate and raises false-positive rate, while prompts induce a strong trade-off between the two.
Our confidence analysis shows that models become more uncertain on multi-event audio, revealing room for improvement.
\end{abstract}

\section{Introduction}
Recently, large audio-language models (LALMs)~\cite{af3, audsemthinker, step2audio, mimoaudio} and omni models~\cite{Qwen2.5, qwen3, gemma3n} have demonstrated strong and versatile reasoning ability over diverse audio inputs.
For real-world usage, it is important to assess a model's faithfulness and robustness on complex audio events, because many wild acoustic scenes are inherently multi-event, making single-event evaluation insufficient.
Prior work~\cite{echo} evaluates hallucinations in LALMs by extracting candidate sound objects from captions via noun parsing and querying their presence. 
While adversarial sampling can create challenging queries, it can also introduce ontology-inconsistent labels (\texttt{Yes} for \textit{bird} but \texttt{No} for \textit{animal}), which penalizes evidence-consistent answers and can be mistaken for hallucinations.
Manual annotation~\cite{ahabench,cmmbench} enables fine-grained hallucination analysis, but the resulting evaluation sets are typically small ($\leq$400 audio samples), limiting coverage over diverse acoustic conditions. 
Meanwhile, recent benchmarks~\cite{mmau,audiotrust,mmar} report task accuracy but do not isolate how auditory scene complexity affects event grounding and false alarms.
To address these limitations, we build a large-scale multi-event evaluation from AudioCapsV2 and analyze four SOTA Audio LLMs' responses. Our contributions are:
\begin{itemize}
    \item We extract and normalize structured $\sim$145K events from AudioCapsV2 (71K clips) to construct a multi-event evaluation set, consisting of 578 audio event descriptions.
    \item We generate $\sim$356K absent-event queries by embedding each event in an audio-aligned text space and sampling absent-events that are semantically distant from present-events.
    \item We quantitatively characterize detection performance as scene complexity grows across 12 prompts, finding that increasing the number of audio events consistently lowers true-positive rates on present-events ($\sim -29\ \mathrm{pp}$) and raises false-positive rates on absent-events ($\sim +8\ \mathrm{pp}$).
    \item We analyze prompt sensitivity, finding a trade-off: prompts that maximize true positives often also increase false alarms. In other words, phrasing that biases the model toward “Yes” will improve recall but worsen hallucinations, and vice versa.
    \item Our analysis shows that as audio becomes more complex, models show less confidence on correct responses and more varied confidence on incorrect responses, \textcolor{black}{revealing that models become more uncertain with increased complexity.}
\end{itemize}
Our code for dataset extraction and evaluation is available on our project page ~\url{https://github.com/alm-evaluation/multi-event}

\section{Building Queries for Event Grounding}

\begin{figure*}[t]
  \centering
  \includegraphics[width=\textwidth]{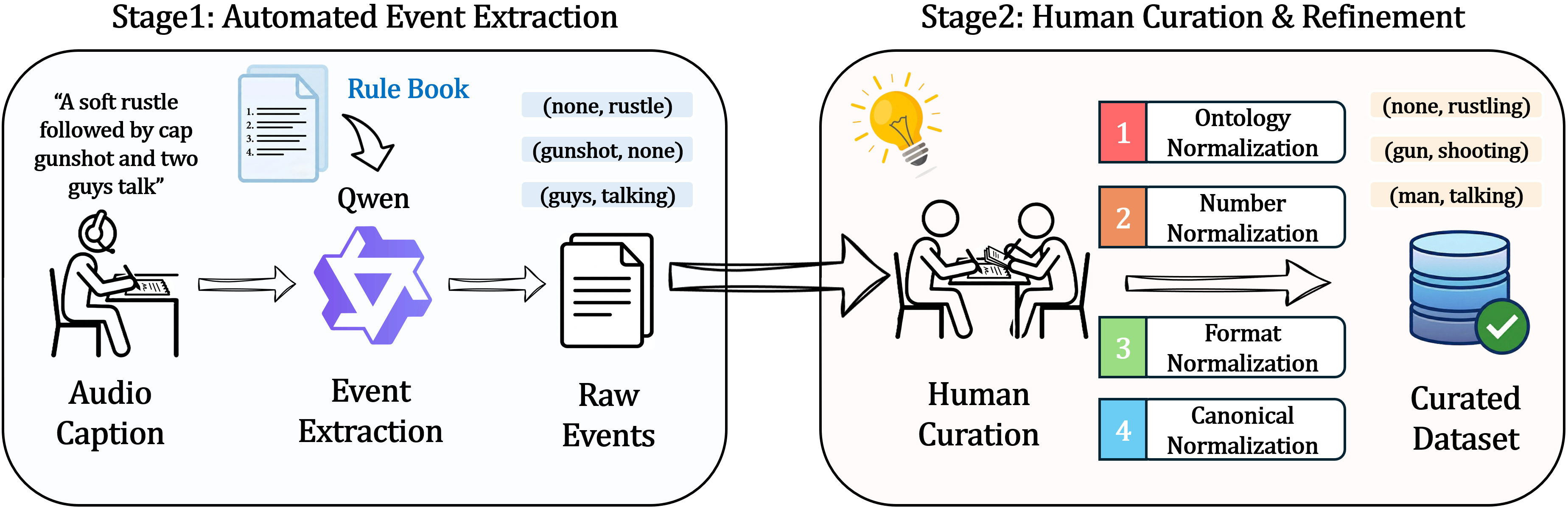}
  \caption{Overall pipeline of audio event extraction.}
  \label{fig:pipeline}
\end{figure*}

\subsection{Audio event extraction}
We use AudioCapsV2~\cite{audiocaps} as our primary source for audio event extraction for three reasons:
(1) it provides $\sim$90K audio clips for large-scale analysis,
(2) captions are written by human annotators and typically describe salient audio events, and
(3) the annotation protocol prioritizes audible content, making captions largely audio-grounded.
These properties make AudioCapsV2 suitable for constructing event and query sets.

For each caption, we extract (\textit{source}, \textit{attribute}) pairs using the Qwen3-Next-80B-A3B-Instruct-FP8 model.
A \textit{source} is a concrete sound producer explicitly mentioned in the caption (\textit{person}, \textit{dog}), and
an \textit{attribute} is an explicitly stated acoustic action or texture (\textit{talking}, \textit{barking}).
To improve consistency and reduce variability in the extracted events, we enforce a strict extraction policy:
(i) extract only events explicitly stated in the caption and ignore uncertain mentions (\textit{maybe});
(ii) treat abstract placeholders or pronouns as invalid sources (\textit{something}) and set $source=\texttt{null}$ when no valid source is present; and
(iii) avoid attributes that merely restate the source (\textit{car sound}) and set $attribute=\texttt{null}$ when no explicit acoustic action or texture term is stated.

However, we find that the model often outputs the same event in different surface forms (\textit{train horn blowing}, \textit{train blowing}).
We therefore first normalize events and then retain only event types with more than 20 occurrences to focus our analysis on common audio events.
Specifically, human annotators manually refine and normalize the extracted events with unanimous agreement according to four principles, consulting a standard English dictionary:
(1) \textbf{Ontology normalization}: consolidating synonymous or hierarchically related sources into unified semantic categories to reduce category sparsity and semantic overlap (\textit{male} $\rightarrow$ \textit{man}, \textit{kitten} $\rightarrow$ \textit{cat});
(2) \textbf{Number normalization}: converting plural sources to their singular forms to ensure consistency in category aggregation and frequency statistics (\textit{people} $\rightarrow$ \textit{person}, \textit{dogs} $\rightarrow$ \textit{dog});
(3) \textbf{Format normalization}: converting verb-form attributes into their gerund (-ing) forms for uniform grammatical representation (\textit{rustle} $\rightarrow$ \textit{rustling}, \textit{whistle} $\rightarrow$ \textit{whistling});
(4) \textbf{Canonical normalization}: rewriting semantically equivalent expressions into a standardized (\textit{source}, \textit{attribute}) form to reduce lexical and morphological variation ($(\textit{man}, \textit{giving speech}) \rightarrow (\textit{man}, \textit{speaking})$, \textit{gunshot} $\rightarrow (\textit{gun}, \textit{shooting})$).

\begin{figure}[ht]
  \centering
  \includegraphics[width=\linewidth]{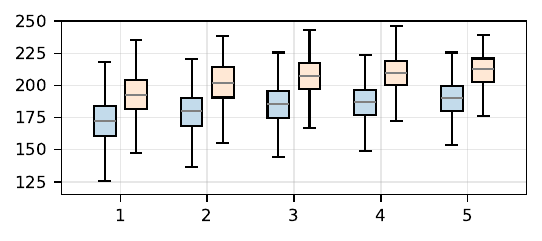}
  \caption{$\operatorname{erank}$ distribution of audio embeddings from EAT (left) and SSLAM (right).}
  \label{fig:erank}
\end{figure}

Through this extraction and normalization pipeline, we obtain 145,236 events from 71,174 audio samples.
Table~\ref{tab:event_distribution} shows that approximately 74\% of the audio samples contain more than one event.
To examine model behavior across a range of auditory complexity, we select audio samples containing at most 5 events.
After filtering and normalization, we obtain $N=578$ unique events $\mathcal{X}=\{x_i\}_{i=1}^{N}$, consisting of 129 sources and 201 attributes.
We denote an audio recording with $n$ events as $A=\{x_1, x_2, \ldots, x_n\}$, where each $x_{i}$ is a normalized (\textit{source}, \textit{attribute}) event.
We evaluate a model's accuracy on the events in each audio sample and refer to this as present-event detection.
To validate that acoustic representations become more complex as event count increases (independently of our extraction process), we compute the effective-rank ($\operatorname{erank}$)~\cite{erank} of embeddings from self-supervised audio models~\cite{eat, sslam}.
Figure~\ref{fig:erank} shows that embeddings tend to exhibit higher $erank$ as the number of events increases.
The instruction used for event extraction is available on our webpage.

\begin{table}[t]
\centering
\caption{Distribution of audio samples by the number of events.}
\begin{tabular}{c|ccccccc}
\hline
\textbf{\# events} & 1 & 2 & 3 & 4 & 5 \\
\hline
\textbf{\# samples} & 25.8\% & 49.3\% & 20.5\% & 3.8\% & 0.5\% \\
\hline
\end{tabular}
\label{tab:event_distribution}
\end{table}

\subsection{Constructing valid absent-events}
To evaluate robustness without inflating the measured false-positive rate, absent-event queries should avoid being \emph{too close} to present-events.
For example, if a clip contains (man, talking), answering \texttt{Yes} to (person, chattering) may be evidence-consistent rather than a hallucination.
Rule-based filters (e.g., WordNet~\cite{wordnet} hypernyms) are often insufficient because acoustic similarity does not always follow lexical similarity (\textit{vibrating} vs.\ \textit{saw running}).
We therefore embed each event $x$ using a text embedding model $f(\cdot)$ with mean-centering normalization and compare events via cosine similarity.
We test general-purpose language models~\cite{qwen3, t5}, but find that they often fail to separate acoustically similar events.
In contrast, ReCLAP~\cite{reclap}, trained with an audio-text contrastive objective, provides better alignment between textual descriptions and acoustic similarity. 
As shown in Table~\ref{tab:embed_compare}, ReCLAP assigns higher similarity to perceptually related pairs, whereas T5 and Qwen3 do not.

\begin{table}[ht]
\centering
\caption{Cosine similarity $\operatorname{sim}(f(x_A), f(x_B))$ and rank of $x_B$ relative to $x_A$ among all events, denoted \#.}
\label{tab:embed_compare}
\fontsize{8}{10}\selectfont
\setlength{\tabcolsep}{2pt}
\begin{tabular}{c|cc|cc}
\toprule
$x_A$ & \multicolumn{2}{c|}{(\textit{water}, \textit{splashing})} & \multicolumn{2}{c}{(\textit{none}, \textit{vibrating})} \\
$x_B$ & (\textit{rain}, \textit{hitting}) & (\textit{oil}, \textit{sizzling}) & (\textit{saw}, \textit{running}) & (\textit{insect}, \textit{buzzing}) \\
\midrule
\textbf{ReCLAP} & 0.59 / \#16 & 0.49 / \#25 & 0.22 / \#95 & 0.17 / \#128 \\
\textbf{T5}  & 0.36 / \#37 & -0.01 / \#230 & -0.11 / \#538 & 0.04 / \#167 \\
\textbf{Qwen3} & 0.33 / \#40 & -0.01 / \#234 & -0.04 / \#321 & 0.14 / \#141 \\
\bottomrule
\end{tabular}
\end{table}
We first consider an audio clip that contains a single event and use cosine similarity to filter semantically similar events with respect to the present-event.
We define the set of dissimilar events for a given event $x$ using a hyperparameter $\alpha$, where $x_{(\alpha)}$ is the $\lfloor\alpha N\rfloor$-th most similar event with respect to $x$ in $\mathcal{X}$.

\begin{equation}
  D_\alpha(x)=\{ x'|\operatorname{sim}(f(x'), f(x)) < \operatorname{sim}(f(x_{(\alpha)}), f(x)) \} \label{eqn:dissimilar}
\end{equation}
In our absent-event detection task, we empirically set $\alpha=0.3$ to balance the diversity of sampled absent-events with sufficient semantic distance from present-events.

\subsection{Audio with multiple events}
\begin{figure}[!ht]
  \centering
  \includegraphics[width=\linewidth]{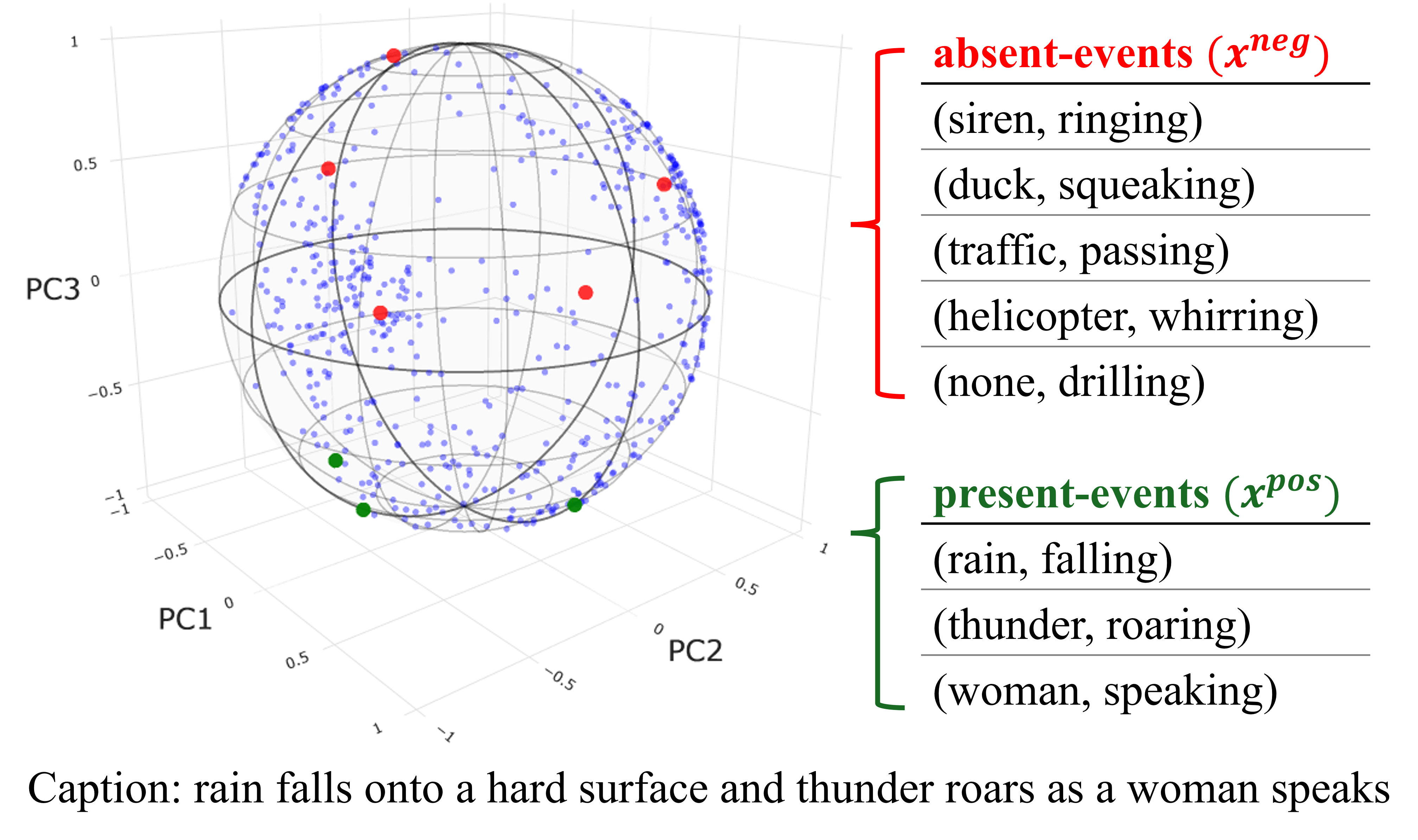}
  \caption{Visualization of the event embedding space on principal component axes. Blue dots indicate all other events.}
  \label{fig:pca}
\end{figure}

As shown in Table~\ref{tab:event_distribution}, most audio recordings consist of multiple events.
Given an audio clip containing $n$ events, we construct a set of absent-events that are distinct from its present-events by taking the intersection, and then sampling $m$ events using a sampling method $\mathcal{S}$:
\begin{gather}
    \{x'_j\}_{j=1}^{m} \sim \mathcal{S}\left(\bigcap_{i=1}^{n} D_{\alpha}(x_{i})\right) \tag{2} \label{eqn:random}
\end{gather}
We use random sampling with $m=5$, resulting in 356K absent-event queries for multi-event clips.
Figure~\ref{fig:pca} visualizes the event embedding space $f(\cdot)$ and shows that the sampled absent-events $(x^{\text{neg}})$ lie far from the present-events $(x^{\text{pos}})$.

\subsection{Audio LLM evaluation setup}

\begin{table}[ht]
\centering
\caption{Question templates and response instructions used in prompts for evaluating the existence of $x$ in an audio sample.}
\label{tab:prompt_response}

\begin{tabular}{cl}
\toprule
\multicolumn{2}{l}{\textbf{Question Templates}} \\
$q_1$ & In this audio, is there any \{$x$\}? \\
$q_2$ & Can you hear a sound of \{$x$\}? \\
$q_3$ & Does this audio contain the sound of \{$x$\}? \\
$q_4$ & Is there any occurrence of \{$x$\}? \\
\midrule
\multicolumn{2}{l}{\textbf{Response Instructions}} \\
$r_1$ & Respond using only one word: yes or no. \\
$r_2$ & Answer with only: yes or no. \\
$r_3$ & Output yes or no only. \\
\bottomrule
\end{tabular}
\end{table}

We use vLLM~\cite{vllm} to accelerate inference during evaluation.
We benchmark four publicly available, vLLM-supported audio-capable LLMs with strong performance on recent audio reasoning benchmarks such as MMAU~\cite{mmau}: Qwen3-Omni-30B-A3B-Instruct~\cite{qwen3}, Qwen2.5-Omni-7B and 3B~\cite{Qwen2.5}, and Audio-Flamingo 3-7B~\cite{af3}.
We design four question templates to account for the substantial prompt sensitivity of LALMs~\cite{speechifeval25, audiobench}.
Since models often produce free-form outputs without explicit instructions, we additionally introduce three response instruction templates.
We denote each prompt as $q_ir_j$, yielding 12 prompts (Table~\ref{tab:prompt_response}).
We generate all responses with greedy decoding and run experiments on an RTX PRO 6000 GPU.

\begin{figure}[!t]
  \centering
  \includegraphics[width=0.87\linewidth]{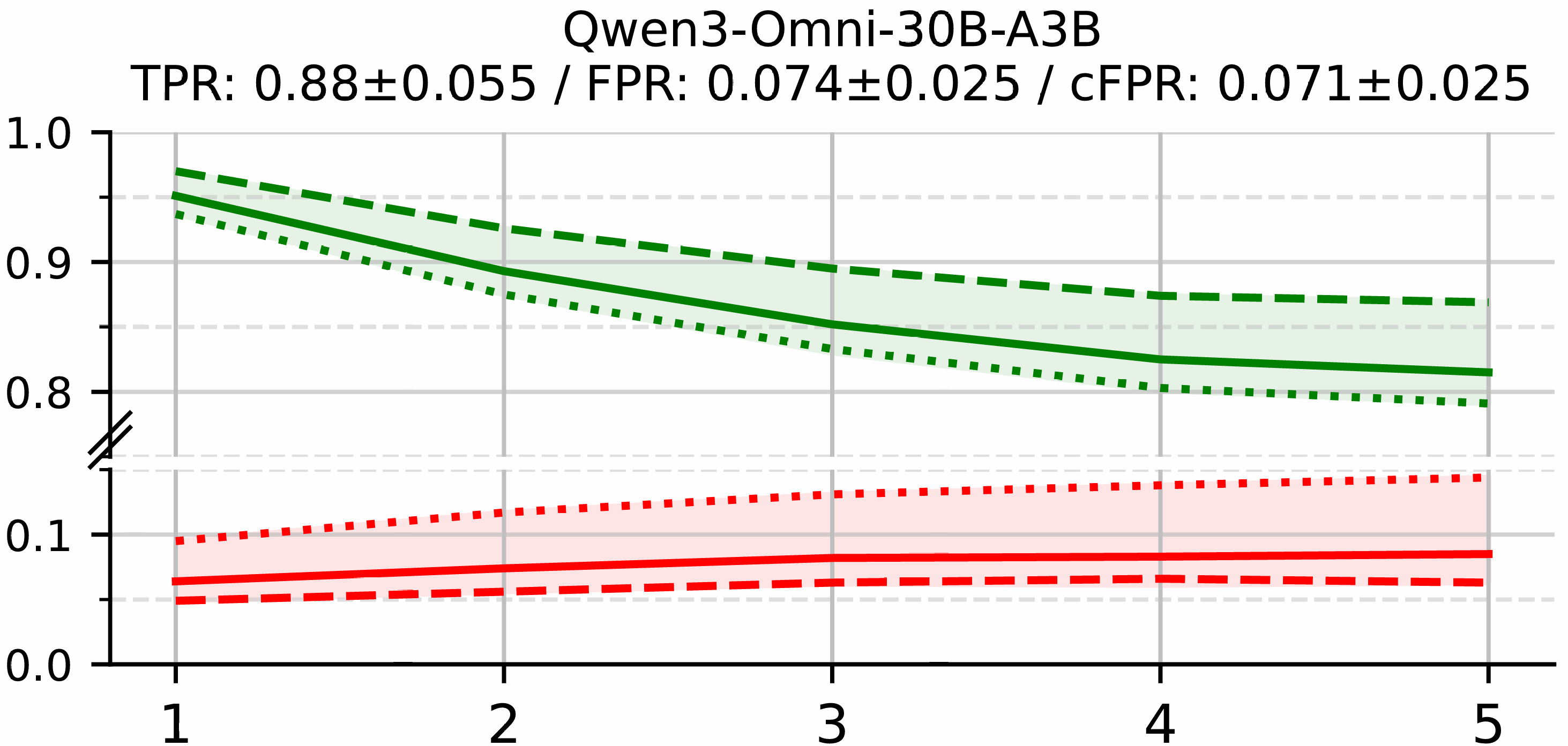}

  \includegraphics[width=0.87\linewidth]{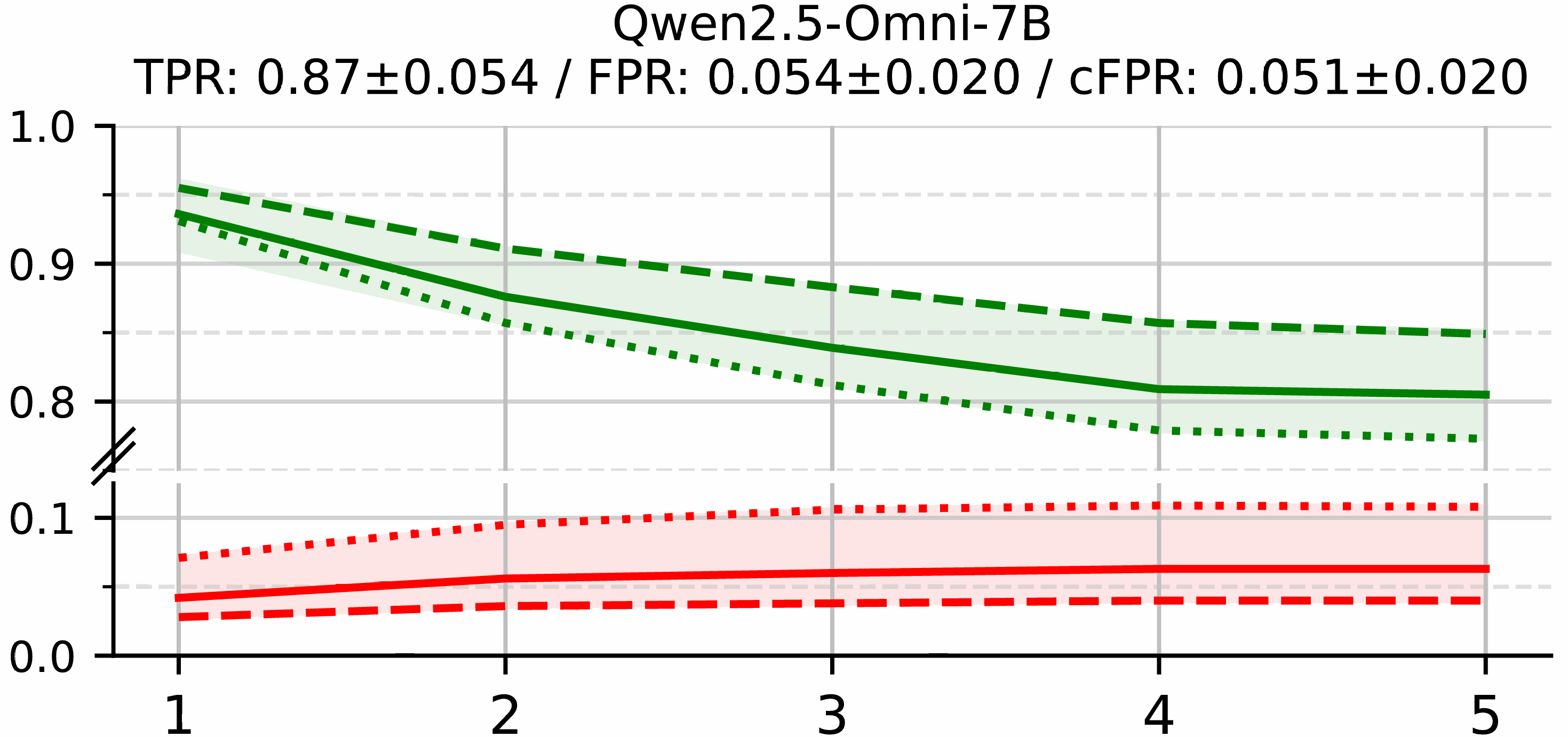}

  \includegraphics[width=0.87\linewidth]{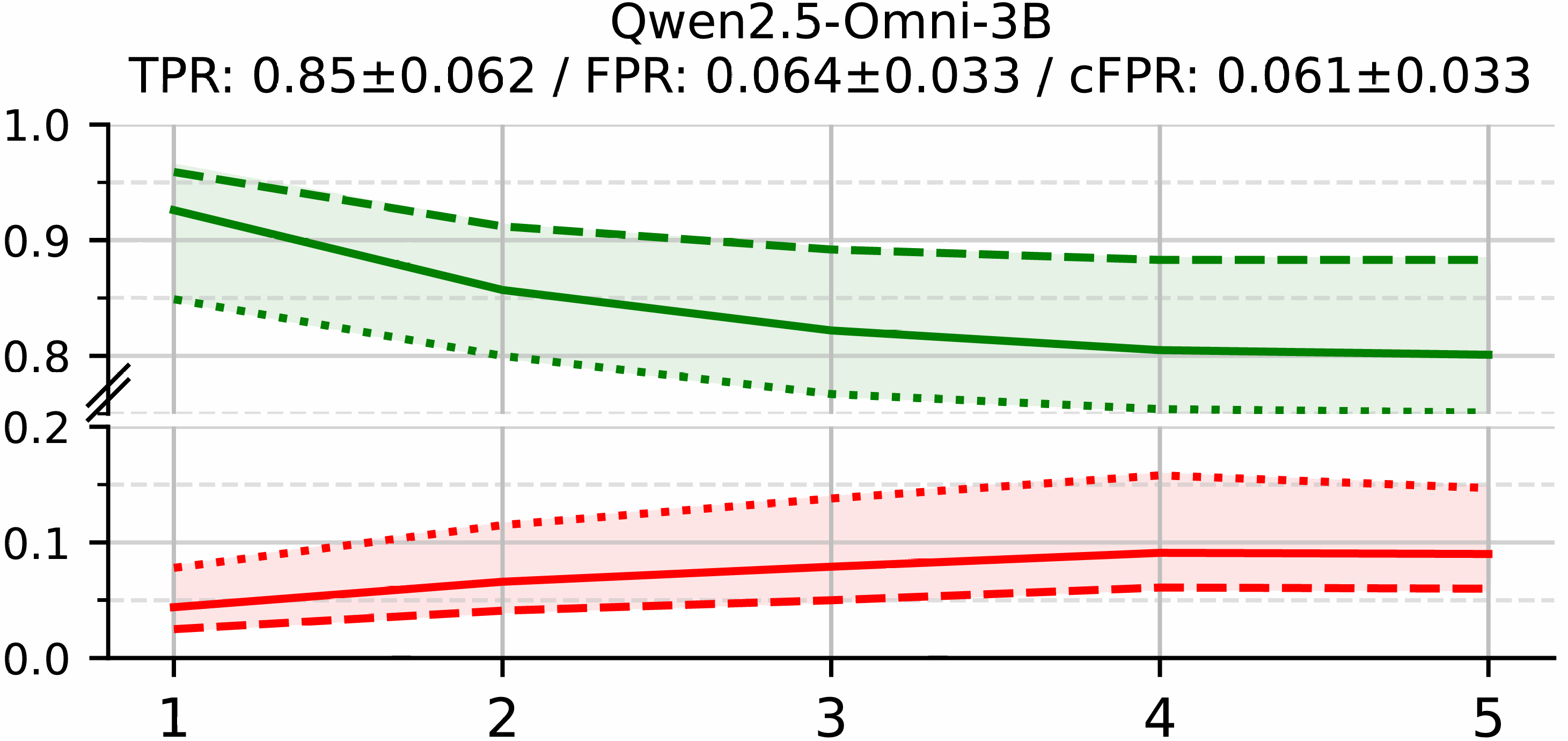}

  \includegraphics[width=0.87\linewidth]{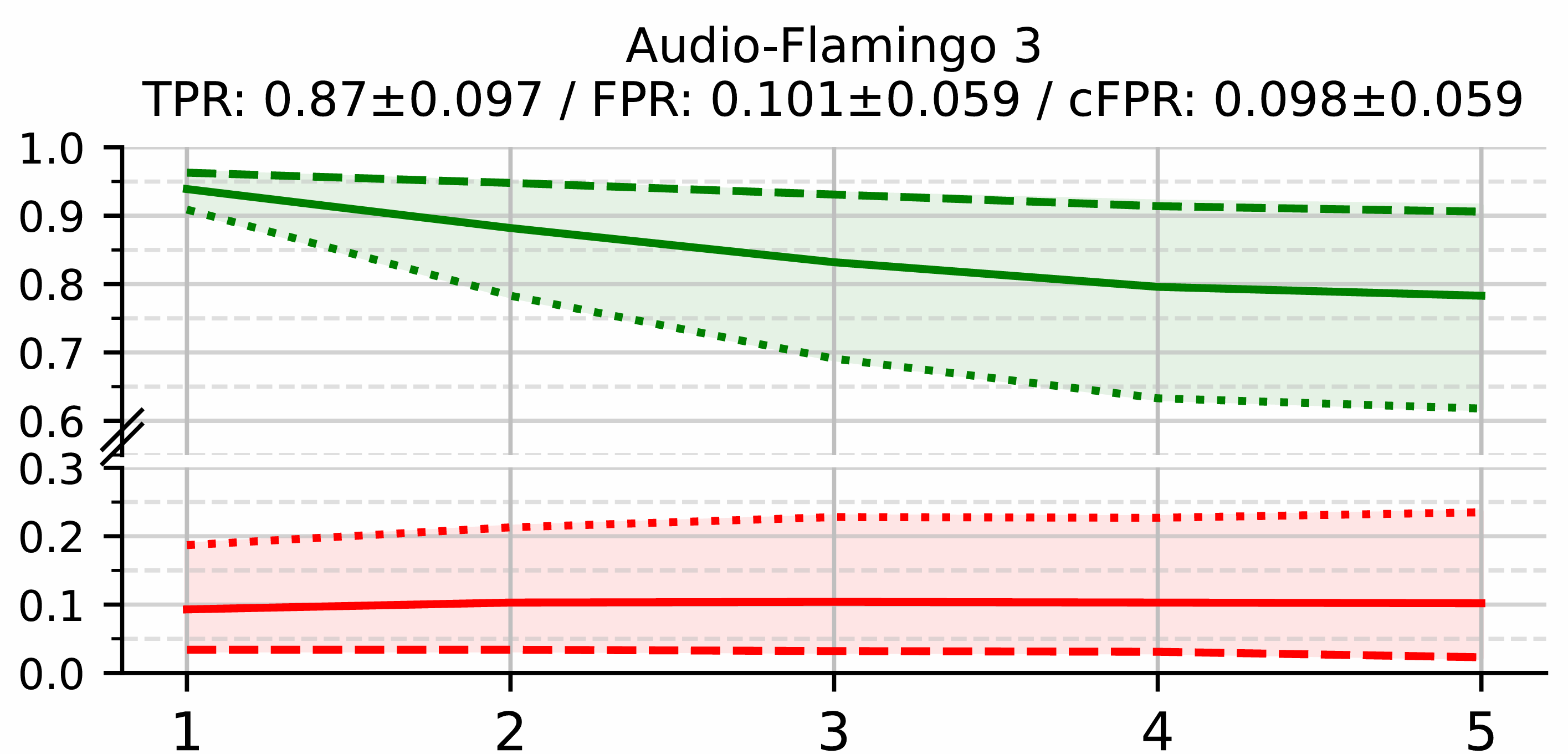}

  \caption{True Positive Rate (green) on present-event detection and False Positive Rate (red) on absent-event detection, plotted against the number of events across audio LLMs. 
  Solid (\textbf{-}), dashed ($--$) and dotted ($..$) lines indicate the average across prompts, the prompt with the highest TPR / lowest FPR, and the prompt with the lowest TPR / highest FPR, respectively.}
  \label{fig:prediction-graphs-1x4}
\end{figure}

\begin{figure*}[!ht]
\centering

\setlength{\tabcolsep}{3pt}     
\renewcommand{\arraystretch}{0} 

\begin{tabular}{@{}cc@{}}

\includegraphics[width=0.49\textwidth]{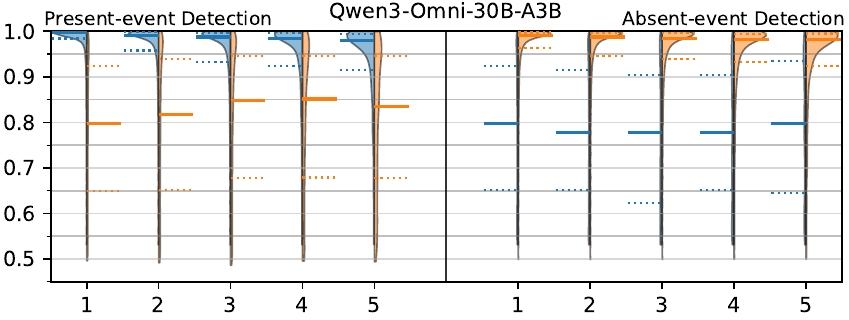} &
\includegraphics[width=0.49\textwidth]{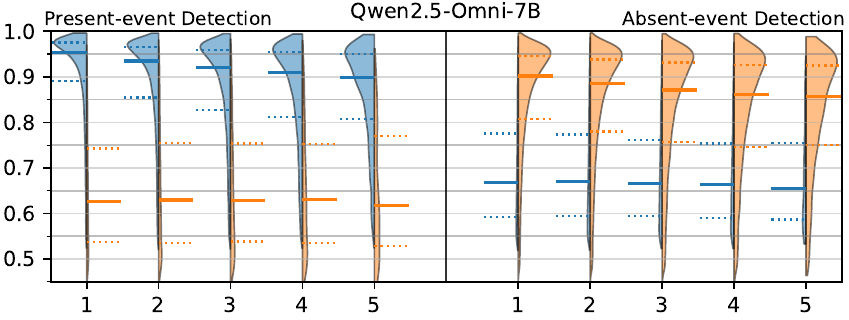} \\

\includegraphics[width=0.49\textwidth]{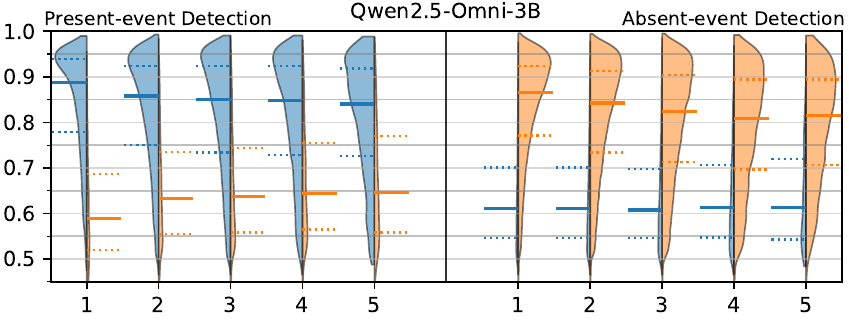} &
\includegraphics[width=0.49\textwidth]{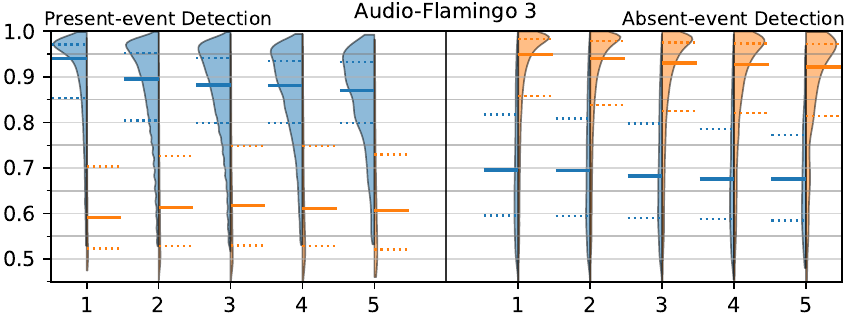} \\

\end{tabular}

\caption{Normalized output token probability distribution of Audio LLMs. \texttt{Yes} / \texttt{No} scores are at the left / right axis, respectively.}
\label{fig:confidence}

\end{figure*}

\begin{table}[ht]
\centering
\caption{Prompts and their corresponding performance (\%).
$\text{PD}_h$ / $\text{PD}_l$ denote the prompts with the highest / lowest TPR on present-event detection.
$\text{AD}_l$ / $\text{AD}_h$ denote the prompts with the lowest / highest FPR on absent-event detection.
}
\label{tab:prompt_tau}
\fontsize{8}{10}\selectfont
\setlength{\tabcolsep}{3pt}
\begin{tabular}{c|cccc}
\toprule
 & $\text{Qwen}_{30B}^3$ & $\text{Qwen}_{7B}^{2.5}$ & $\text{Qwen}_{3B}^{2.5}$ & $\text{AF}_{7B}^{3}$ \\
\midrule
$\text{PD}_h$    & $q_2r_2/91.8$  & $q_1r_1/90.3$  & $q_1r_1/90.9$  & $q_1r_3 /94.2$ \\
$\text{PD}_l$    & $q_4r_3/86.4$  & $q_3r_3/84.6$  & $q_4r_2/79.3$  & $q_3r_3/75.8$ \\
$\text{AD}_l$    & $q_4r_3/5.6$   & $q_3r_2/3.6$   & $q_4r_2/3.9$   & $q_3r_1/3.4$  \\
$\text{AD}_h$    & $q_2r_2/11.5$  & $q_1r_1/9.1$   & $q_1r_1/11.2$  & $q_4r_1/21.0$  \\
\midrule
$\tau_\text{bias}$     & $-0.90$ & $-0.66$ & $-0.87$ & $-0.72$ \\
\midrule
$r$  & $-0.949$ & $-0.791$ & $-0.730$ & $-0.848$ \\
$p$  & $2.41\times10^{-6}$ & $2.18\times10^{-3}$ & $6.97\times10^{-3}$ & $4.87\times10^{-4}$ \\

\bottomrule
\end{tabular}
\end{table}

\section{Experiment Results}

\subsection{Impact of Auditory Complexity on Model Faithfulness}
Figure~\ref{fig:prediction-graphs-1x4} shows that as event count increases, TPR drops in present-event detection and FPR rises in absent-event detection.
We also observe substantial prompt sensitivity.
In particular, Audio-Flamingo 3 shows the largest fluctuations across prompts, whereas Qwen2.5-Omni-3B shows less variation.
We further test whether false alarms decrease when a model correctly identifies most present-events.
Specifically, we compute a \emph{conditional FPR} (cFPR) over clips where the model correctly predicts at least 75\% of the present-events, and then measure the FPR on the corresponding absent-event queries.
Across models, cFPR is at most 0.3 percentage points lower than the overall FPR, suggesting that false alarms are only weakly coupled to present-event recognition.
Overall, these results indicate that current SOTA Audio LLMs still struggle to reliably distinguish \emph{what is present} from \emph{what is absent} in complex acoustic scenes.

\subsection{Prompt effects on present/absent-event detection}
Table~\ref{tab:prompt_tau} summarizes, for each model, the prompts with the highest and lowest TPR on present-event queries, as well as the lowest and highest FPR on absent-event queries.
We observe a clear trade-off: prompts that yield higher TPR often also yield higher FPR, reflecting a stronger tendency to answer \texttt{Yes}; conversely, prompts that reduce FPR tend to also reduce TPR.
This suggests that prompt variations can induce a \texttt{Yes}/\texttt{No} response bias beyond the intended instruction.
To quantify this bias, we compute Kendall’s $\tau$~\cite{kendall} between prompt rankings by TPR (descending) on present-event queries and by FPR (ascending) on absent-event queries, and denote it as $\tau_\text{bias}$. 
Across models, $\tau_\text{bias}$ is consistently negative, indicating that prompts ranked highly for present-event detection tend to rank poorly for absent-event detection.
We further assess whether prompt choice affects robustness to auditory scene complexity.
For each prompt, we compute a complexity gap from 1-event to 5-event clips ($\Delta\mathrm{TPR}=\mathrm{TPR}_{1}-\mathrm{TPR}_{5}$ and $\Delta\mathrm{FPR}=\mathrm{FPR}_{5}-\mathrm{FPR}_{1}$), and then compute the Pearson correlation ($r$) and its p-value ($p$) between $\Delta\mathrm{TPR}$ and $\Delta\mathrm{FPR}$ across prompts.
As shown in Table~\ref{tab:prompt_tau}, the correlation is strongly negative across models ($r=-0.73$ to $-0.95$), revealing a trade-off in complexity sensitivity: prompts that induce larger TPR drops tend to induce smaller increases in FPR, and vice versa.
This highlights that prompt sensitivity to audio complexity in one task tends to come with reduced sensitivity in the other.

\subsection{Model confidence analysis}
To investigate whether models exhibit different degrees of uncertainty when processing more complex audio samples~\cite{mcrbench, audiolens}, we extract each model’s output token probabilities~\cite{uncertainty_quantification} and compute the normalized probability mass over the answer set, as defined in Equation~\ref{eqn:conf}.

\begin{gather}
  \operatorname{conf}(y,x)=\frac{p(y|x)}{p(\texttt{Yes}|x)+p(\texttt{No}|x)},\quad y\in\{\texttt{Yes}, \texttt{No}\} \tag{3} \label{eqn:conf}  
\end{gather}
Figure~\ref{fig:confidence} shows that confidence scores for correct outputs ($\operatorname{conf}(\texttt{Yes}, x^{\text{pos}})$, $\operatorname{conf}(\texttt{No}, x^{\text{neg}})$) exhibit long-tailed distributions; whereas confidence scores for incorrect outputs ($\operatorname{conf}(\texttt{No}, x^{\text{pos}})$, $\operatorname{conf}(\texttt{Yes}, x^{\text{neg}})$) are comparatively flat.
As the number of events in an audio clip increases, confidence for correct outputs decreases.
In contrast, confidence for incorrect outputs shows no consistent trend across models.
Overall, these results suggest that increased auditory complexity amplifies a \texttt{Yes} response bias and reduces confidence in correct \texttt{No} decisions, indicating increased uncertainty on absent-event judgments in complex acoustic scenes.

\section{Conclusion and Discussion}
We present a large-scale sensitivity analysis of multi-event audio grounding in audio-capable LLMs using present-event and similarity-filtered absent-event queries to measure how scene complexity affects grounding and false alarms.
Across four SOTA audio LLMs, increasing event count consistently reduces TPR and raises FPR, indicating that separating present from absent events remains challenging in complex scenes.
Prompt choice strongly affects robustness and often trades higher recall for more false alarms.
Our confidence analysis further suggests increased uncertainty in complex audio, highlighting that current audio LLMs have substantial room for improvement in handling complex scenes.
We hope our analysis will facilitate future research on faithful audio grounding.

\section{Generative AI Use Disclosure}
We employed generative AI tool (ChatGPT) exclusively to support English proofreading and improve language quality. No scientific content, experimental results, or conclusions were generated by the tool. All authors carefully reviewed and incorporated edits as appropriate and remain fully responsible for the content of this paper and its submission.

\bibliographystyle{IEEEtran}
\bibliography{mybib}

\end{document}